\documentclass[floatfix,prl,superscriptaddress,twocolumn,english,showpacs,letterpaper]{revtex4}
\usepackage{graphicx}
\usepackage{amsmath}
\usepackage{amssymb}
\usepackage{color}
\usepackage{wasysym}
\makeatletter

\begin{document}

\title{Zero and low temperature behavior of the two-dimensional $\pm J$ Ising
spin glass}
\author{Creighton K. Thomas}
\affiliation{Department of Physics and Astronomy, Texas A\&M University,
College Station, Texas 77843-4242, USA}
\author{David A. Huse}
\affiliation{Department of Physics, Princeton University, Princeton, NJ 08544, USA}
\author{A. Alan Middleton}
\affiliation{Department of Physics, Syracuse University, Syracuse, NY 13244, USA}
\begin{abstract}
Scaling arguments and precise simulations are used to study the
square lattice $\pm J$ Ising spin glass, a prototypical model for glassy systems.
Droplet theory explains, and our numerical results show,
entropically-stabilized long range spin glass order
at zero temperature, which resembles the energetic
stabilization of long range order in higher dimensional models at finite temperature.
At low temperature, a temperature-dependent
crossover length scale is used to predict the power-law dependence on temperature of the
heat capacity and clarify the importance of disorder distributions.
\end{abstract}
\maketitle

Glassy systems, characterized by extremely slow relaxation and
resultant complex hysteresis and memory effects, are difficult to
study because their dynamics encompass a great range of time scales
\cite{REVIEW}. Glassy materials include those without intrinsic
disorder, such as silica glass, and those where quenched disorder
influences the active degrees of freedom. An example of
the latter is the Edwards-Anderson spin glass model \cite{EA},
which includes the disorder and frustration necessary to capture
many of the complex behaviors seen in disordered magnetic materials.
Many aspects of this prototypical spin glass model 
remain
poorly understood.  The droplet and replica-symmetry-breaking pictures
provide distinct views of spin glasses \cite{YoungBook}.
Analytical results are few, so numerical approaches are invaluable for both
testing and motivating new ideas.
But numerics can also be exceedingly difficult in the general case:
computing spin glass ground states is
believed to require
exponential time to solve exactly, as it is
an NP-hard problem \cite{Barahona,computerScience}.

A fortunate special case that avoids this computational
intractability is the two-dimensional Ising spin glass (2DISG).  The
Hamiltonian is $\mathcal{H} = \sum_{\langle ij \rangle}
J_{ij} s_{i} s_{j}$. 
The nearest-neighbor
couplings ${\mathcal J}=\{J_{ij}\}$ are independent
random variables coupling Ising spins $s_i=\pm 1$ at
sites $i$ 
on a
square lattice with $L^2$ sites.
The random 
signs of the $J_{ij}$ lead to competing interactions
that can not be simultaneously satisfied. 
Here we study the model with the $\pm J$
distribution, where each bond value is $J_{ij}=\pm 1$ with equal
probability.
Highly developed numerical algorithms
\cite{Barahona,GSKastCities,SaulKardar,GLV,ThomasMiddletonSampling}
can efficiently circumvent the complexity due to disorder and frustration:
ground states and finite-temperature partition functions of the 2DISG may be
computed in time polynomial in $L$.
These algorithms have given us more insight
into model glassy systems.

In this Letter, we develop the droplet scaling theory as applied to the
zero and low temperature properties of the $\pm J$ model.
We obtain reliable numerical results for very low temperatures
$T$ and large $L$ (well beyond any Monte Carlo simulation results), including the probability
distributions of spin correlations.
We show that the $\pm J$ model at $T=0$ in 2D has equilibrium
correlations similar to those of a higher-dimensional spin glass
at $T>0$ in its spin glass ordered phase.
Thus the 2DISG provides a computationally-tractable model with spin-glass order that closely
mimics spin glass order in higher dimensions, where large samples can not be
studied numerically.

The ground state of a $\pm J$ model is highly degenerate with an extensive
entropy \cite{vt,mb,SaulKardar}.  For many years it was
assumed that the ground state is critical, with power-law spin-spin correlations
as a function of distance \cite{mb,by,PB,hartmann}.
Jorg, {\it et al.} \cite{Jorg-etal} then presented evidence that instead the ground state has true
long-range spin glass order.  Recent results \cite{roma} about rigid clusters of spins with fixed
relative orientations are also suggestive of long range correlations.
Here we add to that evidence, and present the
corresponding droplet theory, which differs in some important respects \cite{thm} from the
scenarios proposed in Refs. \cite{Jorg-etal,roma,tpv}.
We show how the $T=0$ properties
predict a new temperature-dependent crossover
length scale $\ell_x(T)$ where the probability
distribution of the droplet free energies 
crosses over from
discrete to continuous.  We use this to predict
the low-$T$ scaling of the specific heat.  
This
prediction is shown to agree well with our numerical results, resolving the
long-debated question of the specific heat behavior in this
model.
These results provide new insight into the role of entropy and the importance of the choice of disorder
distribution in spin glasses.

{\bf Computational technique} We start by computing the partition
function $Z_{\mathrm P}({\mathcal J})$ in an $L$-by-$L$ sample with periodic boundary
conditions, adopting previously published techniques \cite{SaulKardar,GLV}, but using
arbitrary precision arithmetic \cite{ThomasMiddletonSampling}.
The efficient computation of $Z$ relies on a hierarchical decomposition of the sample \cite{GLV}:
starting from the smallest
pieces of the spatial decomposition, single plaquettes, neighboring pieces are joined together to recursively
compute the partition function.
The same four Pfaffians \cite{ThomasMiddletonSampling} used to find $Z_{\mathrm{P}}$ also give
$Z_{\mathrm{AP}}$, the partition function for antiperiodic boundary
conditions, where the horizontal
bonds along a vertical column have negated $J_{ij}$.
Note that these computations are exact to within
numerical precision: there is no
question of convergence as there is with Markov chain Monte Carlo methods and we have verified the
numerical stability of our procedure \cite{ThomasMiddletonSampling}.
The two partition functions differ due to a domain wall
induced by the change in boundary conditions.
We use finite differences over $T$ of $F=-kT\ln(Z)$
to compute equilibrium averages for energies
$E$ and entropies $S$.
The sample-dependent differences
$\delta X({\mathcal J})\equiv X_{\mathrm{P}}-X_{\mathrm{AP}}$, with $X=F$, $E$, and $S$,
give relative domain wall (free) energies and entropies.
Error bars in our plots indicate $1\sigma$ statistical errors.

\begin{figure}[h]
\centering
(a)\includegraphics[width=1.5in]{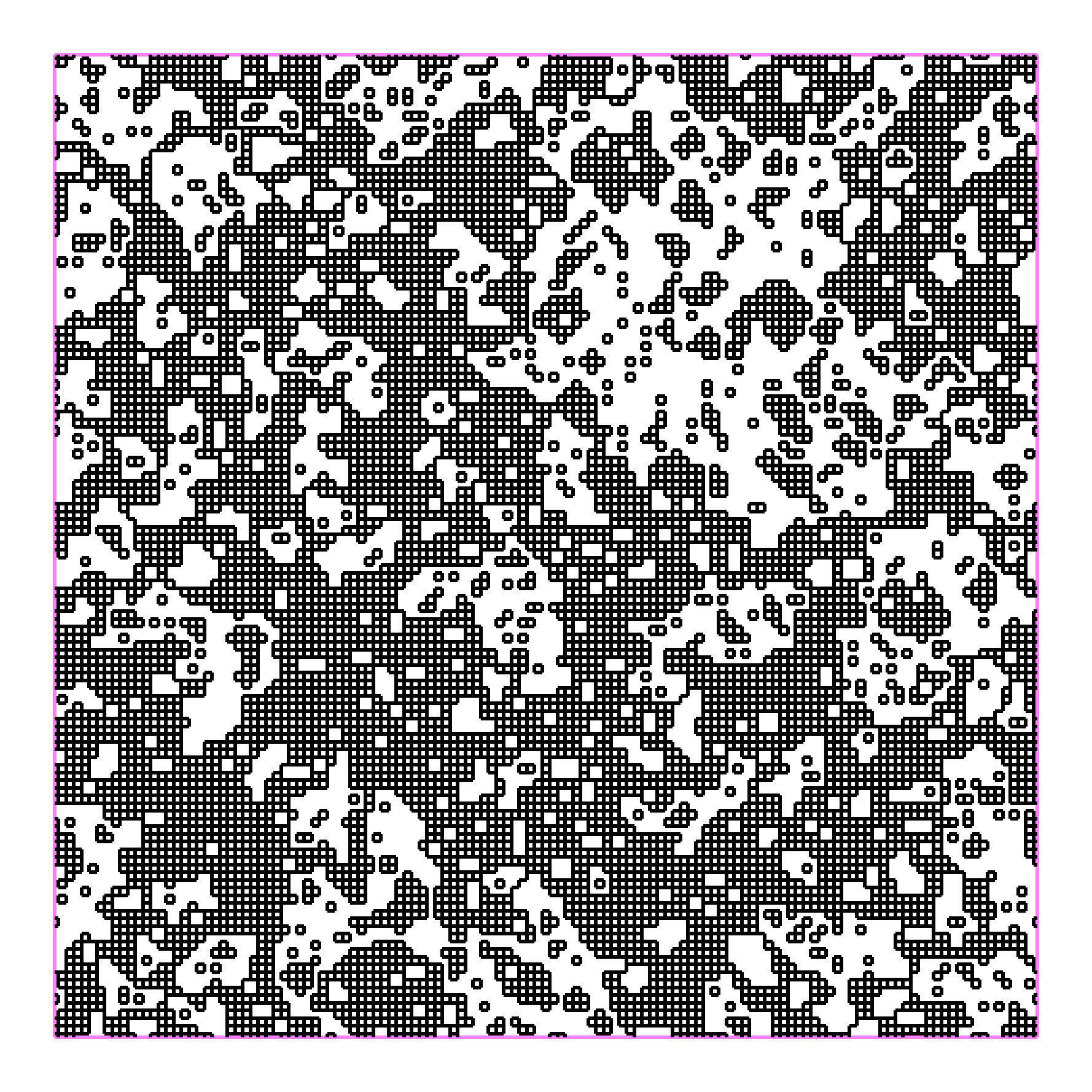}
(b)\includegraphics[width=1.5in]{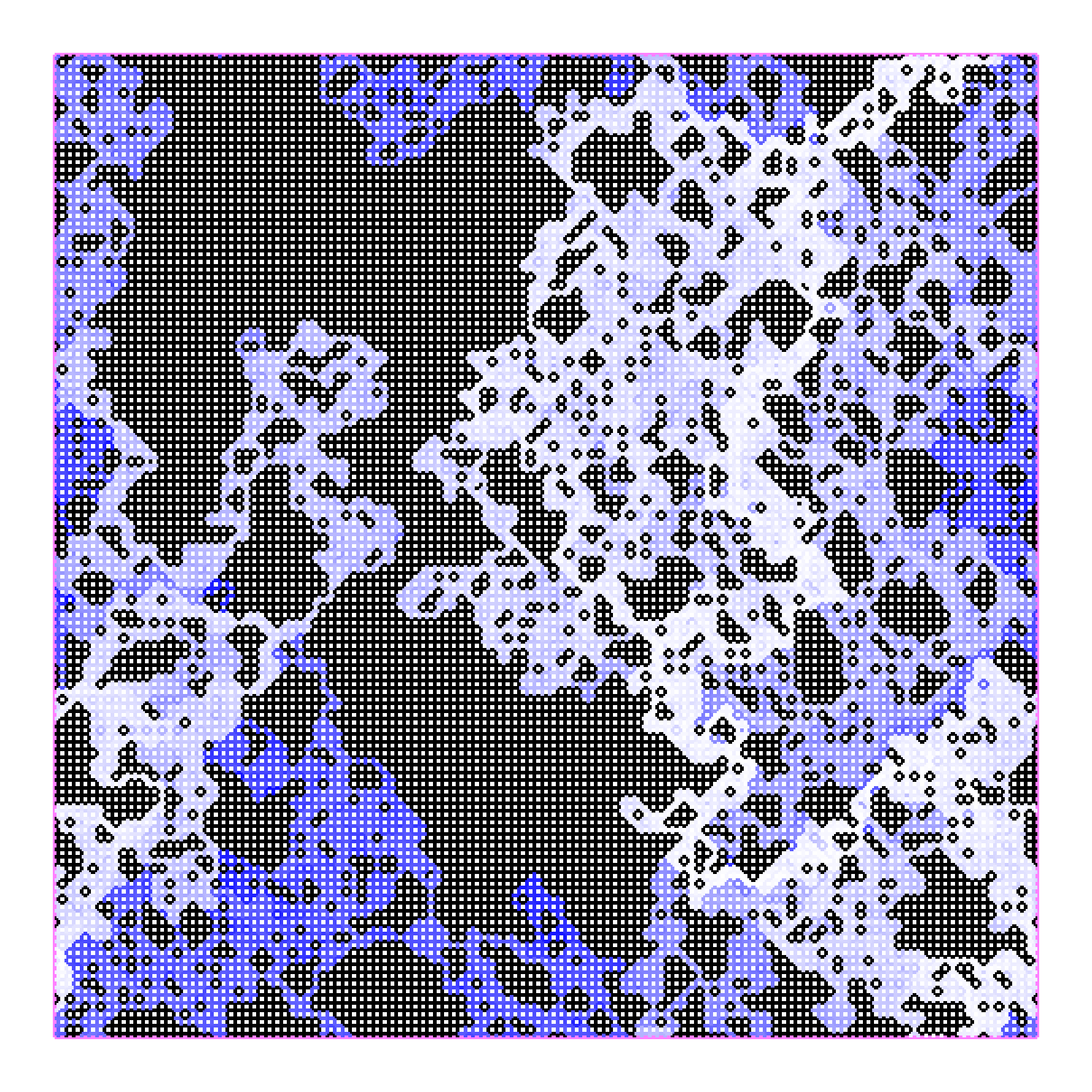}
\caption{ [color online]
An $L=128$ sample.
(a) Nearest neighbor correlations: Black lines are bonds
that are rigidly
correlated at $T=0$.
(b) Relative domain wall. Black lines are bonds whose correlations are not affected by the change $\mathrm{P}\rightarrow\mathrm{AP}$
in the horizontal boundary conditions.  The shaded bonds indicate the change
in the correlations on a logarithmic scale, with light colors for bonds with $|\delta\langle s_is_j\rangle_0|\sim 1$,
and the darkest colors indicating $|\delta\langle s_is_j\rangle_0|\sim e^{-20}$.
This ``diffuse'' domain wall appears to have a multifractal geometry \cite{next}.
}
\label{fig:sample}
\end{figure}

We have extended this technique to 
compute arbitrary spin-spin correlation functions
\cite{cktaamUnpub}.
These correlations can be computed at sufficiently low $T$,
$T=0.02$ for $L\le 256$, and with sufficient precision, $4\,096$ bits,
such that the contribution of excited states can be clearly separated from
the ground state contribution to correlations. If spins have fixed relative orientation with probability greater
than $1-e^{-2/T}$, we take the spins to be rigidly correlated at $T=0$. Such spins have fluctuations
that are clearly smaller than non-rigid spins by a factor of at least $\approx 10^{21}$, exceeding
multiplicity (entropic) factors for excited states. We have confirmed that the
assignment of rigid correlations is unchanged
if $T$ is lowered or the precision increased, using over 200 samples for $L=128$.
An example of $T=0$ rigid and non-rigid nearest-neighbor
spin correlations $\langle s_is_j\rangle_0$, and the domain wall
due to a change in boundary conditions are shown in Fig.\ \ref{fig:sample}.

{\bf Domain walls, droplets, and rigid clusters}
Consider a given ground state configuration of a
particular realization ${\mathcal J}$ of the $\pm J$ model.
A relative domain wall loop is the boundary of a simply
connected region where all the spins are flipped
relative to the given ground state configuration.
The droplet excitation \cite{FisherHuse} at site $i$ and length scale $\ell$
is the set of all lowest-energy domain wall loops of linear
size of order $\ell$ enclosing site $i$ and of order $\ell^2$ other sites.
We can ask about the probability distributions of the
energy and of
the entropy of this 
droplet excitation.
Here we do not directly measure these droplets, but computations
for the very similar system-size relative domain walls
due to changing between $\mathrm{P}$ and $\mathrm{AP}$ boundary conditions give insight into the
effect of these droplets on the spin-spin correlations.

We first note previous work (e.g., \cite{HY}) which shows that system-size domain walls have
a probability distribution for $\delta E$ that is independent of $L$, as $L\rightarrow\infty$
(given a parity for $L$).
For large even $L$, we find that $77.5 \pm1.0\%$ 
of the samples do have zero energy domain walls.
Thus, assuming the droplets have a similar probability distribution,
the event that there is no zero energy droplet of scale $\ell$ surrounding a site $i$ has probability
strictly between zero and one.
This is also true for smaller scales $\ell/2$, $\ell/4$, etc., with
the correlations between the events at each scale presumably falling off rapidly
across 
scales.  As a result \cite{hartmann},
a given site $i$ has no zero energy droplets at any scale less than $\ell$ with a
probability $p(\ell)$ that decreases with increasing $\ell$ as a power of $\ell$.

In a finite sample at $T=0$, we choose to identify the ``backbone'' of the
ground states as the largest set of 
rigidly correlated spins \cite{roma}.
Spins on the backbone
can not be flipped at zero energy by any droplets without flipping the entire backbone.
By the above argument, the number of such spins
scales with a power of $L$ less than two: 
the backbone is a fractal.
Hartmann's results imply a backbone fractal dimension $d_b=1.78(2)$ \cite{hartmann};
our results for
rigidly correlated clusters are consistent with this \cite{next}.
As seen in previous work \cite{roma}, we sometimes see backbone spins
that are not connected by a path of rigid bonds; the backbone can have ``gaps''.
Thus the backbone of the sample in Fig.~1(a)
is all spins on the largest cluster connected by rigid bonds plus
possibly some other spins or clusters whose rigidity is not detected by 
nearest-neighbor correlations.
The backbone on this sample percolates; the fraction of large samples
where the backbone percolates is a number between zero and one \cite{next}.

{\bf Entropic enhancement of correlations}
So far, we have discussed the rigid correlations that arise from locations
where there are no zero energy droplets.  What about zero
energy droplets, where spin clusters flip in the ground state?
Saul and Kardar \cite{SaulKardar} found that zero-energy relative domains walls
have a typical entropy $\delta S \sim L^{\theta_S}$, with $\theta_S\cong 0.50$.
Our data from $L=32$ to $L=256$ is very well fit by a power law, with $\theta_S=0.50\pm 0.01$,
(in contrast with Ref.\ \onlinecite{AP}).
It is intriguing that this exponent is consistent with the simple rational number 1/2.
The probability distribution of $\delta S$ is continuous through zero, so
$|\delta S|$ is less than or of order one with a probability $\sim L^{-\theta_S}$.
We take the probability distribution for $\delta S$ for the zero-energy droplets
to have the same scaling.

This leads to an essential new aspect of the droplet picture for the
two-dimensional $\pm J$ Ising spin glass at $T=0$:  A droplet at
a large $\ell$
typically either has
a nonzero energy so never flips or has a large entropy so it is flipped
away from its usual orientation only with exponentially small
probability $\sim \exp{(-\ell^{\theta_S})}$.  It is only the {\it entropically active}
droplets with $|\delta S|$ less than or of order one that flip at $T=0$ with a
substantial probability.  These active droplets are power-law rare in the length
scale $\ell$: only a fraction $\sim \ell^{-\theta_S}$ of the sample's area is active
at scale $\ell$.  This scenario is very similar to the expected behavior within the
droplet theory \cite{FisherHuse} for an Ising spin glass in 3 or
more dimensions at a nonzero temperature within the
spin glass ordered phase, where only a fraction $\sim \ell^{-\theta}$ of the droplets at scale $\ell$
have $\delta F$ less than order of order $k_BT$ so
are {\it thermally} active, and the typical free energy of a droplet scales
as $\delta F \sim \ell^{\theta}$, with $\theta>0$.

This droplet picture allows us to apply many of the predictions of the droplet
theory \cite{FisherHuse}, with the usual free energy exponent
$\theta$ replaced by the entropy exponent $\theta_S\cong 0.50$.
In particular, it predicts long range
spin glass order, with the $T=0$ spin-glass correlation function for spins at spacing $\vec{r}$,
$G_0(\vec{r})=[\langle s_{\vec{0}}s_{\vec{r}}\rangle_0^2]$,
where the square brackets denote an average over samples, behaving at large $r$ as
\begin{equation}
G_0(\vec{r})-G_0(\infty)\sim r^{-\theta_S}~.
\end{equation}
There is long range order because the total probability, found by summing over all scales,
of flipping a given spin is bounded away from unity, since the probability of the droplet
being entropically active decays as $\ell^{-\theta_S}$.  The active droplets with scale
$\ell>r$ are unlikely to flip only one of a pair of a spins separated by distance $r$, so such droplets do not
substantially reduce $G_0(\vec{r})$.  Thus $G_0(\vec{r})$ is larger
than $G_0(\infty)$ by an amount proportional to the probability $\sim r^{-\theta_S}$
of droplets of scale $r$ being active.

\begin{figure}[h]
\centering
\includegraphics[width=3.4in]{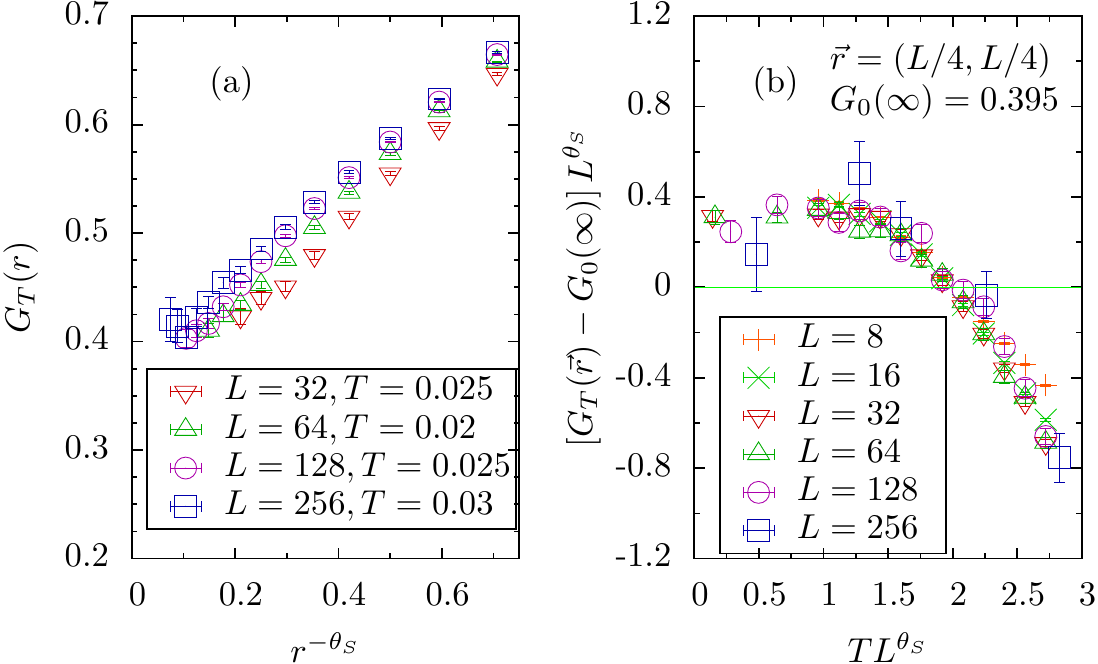}
\caption{(a) Correlation functions at low temperatures for $L=32,64,128,256$ plotted vs.\ $r^{-\theta_S}$,
using $\theta_S=0.50$.
Droplet theory based on the entropy of the droplets predicts a linear curve at small values of $r^{-\theta_S}$ and
large $L$, as is seen.
(b) Sample scaling collapse for correlation function $G_T(\vec{r})$.  Using nonzero-temperature scaling, we expect
$L^{\theta_S}[G_T(\vec{r})-G_0(\infty)]\approx g(\vec{r}/L,TL^{\theta_S})$.  Adjusting $G_0(\infty)$ for the best collapse,
we find $G_0(\infty)=0.395\pm0.010$.
}
\label{fig:corrs}
\end{figure}

The results of numerical calculations for $G_T(\vec{r})$ for $T\ll L^{-\theta_S}$,
shown in Fig.\ 2(a), are consistent with Eq. (1).
To verify this and to more precisely determine
the long range order $G_0(\infty)$,
we measure $G_T(\vec{r})$ for higher $T$.
At nonzero temperature and finite $L$,
the difference between
$G_T(\vec{r})$ and $G_0(\infty)$ should scale as $L^{-\theta_S}g(\vec{r}/L,TL^{\theta_S})$,
for a scaling function $g$.
The scaling parameter $TL^{\theta_S}$ naturally arises from the competition between energy and entropy.
From such scaling of the correlations for several values of $\vec{r}/L$,
we infer $G_0(\infty)=0.395\pm0.010$.
A specific example of a scaling collapse, with fixed $\theta_S=0.50$ and variable $G_0(\infty)$,
is shown in Fig.\ \ref{fig:corrs}(b) for $\vec{r}=(L/4,L/4)$.  In fact this
long range order with $G_0(\infty)\cong 0.4$ is apparent in
Poulter and Blackman \cite{PB} (their Fig. 6), although that is not how they interpreted
those data.
A similar entropically-stabilized long range order has
recently also been seen numerically in a diluted $\pm J$ spin
glass in 3D where the ground state stiffness is zero
\cite{AngeliniRicciTersenghi2011}.

{\bf Crossover to effectively continuous disorder}
Using this droplet picture, we can also understand the crossover to the $T>0$ behavior.
The competition between entropy and energy \cite{BrayMoore, FisherHuse, abm,thm} in spin glasses
causes extreme sensitivity of the long-distance spin
correlations to changes in $T$ (``chaos'').
We use a related argument at very low $T$ to derive 
the low-$T$ scaling of the specific heat in the $\pm J$ model.

At $T=0$, the droplet energies $\delta E$ are integer
multiples of 4, so the distribution of droplet
energies is discrete.  However, the droplet entropies $\delta S(T=0)\sim\ell^{\theta_S}$
are continuously distributed for large $\ell$.
We define a crossover length $\ell_x(T)$ to be the scale
at which the typical $T\delta S(\ell_x)$ becomes $\mathcal{O}(1)$,
causing $\delta F=\delta E-T\delta S$ to become continuously distributed.
Thus
at low $T$
\begin{equation}
\ell_x(T) \sim T^{-1/\theta_S}\,.
\end{equation}
For $\ell\ll\ell_x(T)$ and very low $T$ only zero-energy droplets can be active.
But at scales at and above $\ell_x(T)$ the droplets with $\delta E=4$ or more
can, due to their large entropy, have $\delta F$ near zero and thus also be thermally active
(Ref.\ \onlinecite{Jorg-etal} gave a discussion of this crossover from discrete to continuous
behavior, but not of $\ell_x(T)$.)

\begin{figure}[h]
\centering
\includegraphics[width=3.3in]{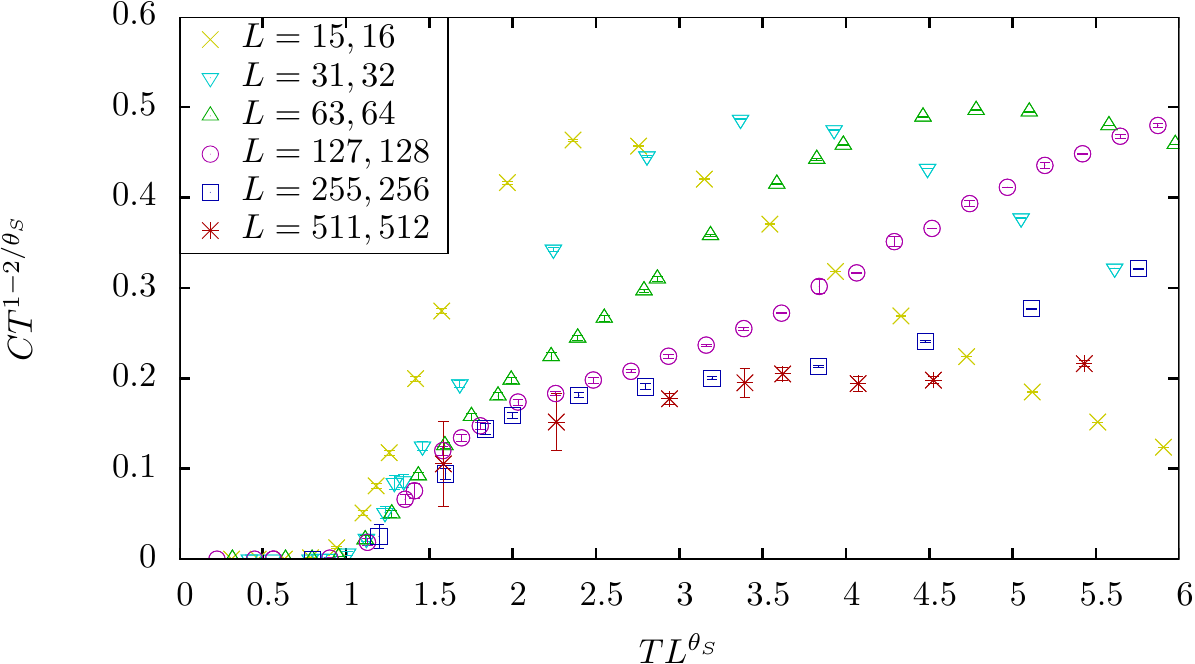}
\caption{ [color online] 
Scaling plot for the low-$T$ specific heat $C$. 
For $T<L^{\theta_S}$, there are no active droplets, so $C$
is exponentially suppressed.  The low-$T$ scaling behavior in large-enough
systems is seen when $C T^{1-2/\theta_S}$ is flat at $TL^{\theta_S}>1$.
The value \cite{LukicEtAlJStat} of $\theta_S=0.50$ is
independently set by the scaling of the entropy of domain walls.
}
\label{fig:specheatPMJ}
\end{figure}

{\bf Specific heat}
The low-$T$ specific heat $C$ 
in a 2DISG is
governed by the thermally active droplets (those with excitation
free energies $\delta F \leq \mathcal{O}(T)$) that have nonzero energy
\cite{FisherHuse}, with the {\it smallest} droplets  
dominating, since they have the highest density.  If the disorder distribution is continuous,
these smallest droplets have size $\mathcal{O}(1)$, 
energy of order $T$ and density proportional to $T$, so
give $C \sim T$ at low $T$.
For $\pm J$ disorder, on the other hand, the droplets with size $\ell<\ell_x(T)$
have either zero energy $\delta E$ or $\delta E = \mathcal{O}(1)$ so 
do not contribute to the low-$T$ specific heat.  The smallest active droplets with 
$\delta E\neq 0$ are of size of order 
$\ell_x(T)$.
Due to their $\mathcal{O}(1)$ energy gap, these droplets each contribute $\sim 1/T^2$ to $C$.
They occur with density $\sim T / \ell_x^2(T)$ so the specific heat
$C\sim T^{(2/\theta_S)-1}$ at low $T$ in an infinite sample (alternatively, the
entropy density is $\ell_x^{\theta_S-2}$, which scales as $C$).
For finite $L$, this power-law specific heat is cut off
at the lowest temperatures when the size $\ell_x(T)$ of
these active droplets exceeds $L$.  This produces the low-$T$ finite-size scaling
form $C\approx T^{(2/\theta_S)-1}c(TL^{\theta_S})$,
where $c(x)$ is a scaling function that goes to a constant for large
argument 
(where $1 \ll \ell_x(T) \ll L$).
Our specific heat data are
shown in this scaling form in Fig. \ref{fig:specheatPMJ}.
Our data for intermediate temperatures, $T \cong 0.35$,
are consistent with Monte Carlo results \cite{KatzgraberLeeCampbell}, which saw an
effective exponent $\alpha\cong -4.2$, with $C\sim T^{-\alpha}$. However, the effective exponent crosses
over to a lower magnitude consistent with
$\alpha = 1-\frac{2}{\theta_S} \cong -3.0$ at low $T$ and the largest $L$.
Contributions from the {\it largest} active droplets give a subleading
correction, which from standard hyperscaling is $\alpha'=-2\nu$ (though see Ref.\ \onlinecite{thm}),
that is too weak to detect yet.
Our result, $\alpha = 1-\frac{2}{\theta_S}$, gives the leading contribution to $C$ at low $T$ 
for the 2D $\pm J$ model in the infinite volume limit.

{\bf Discussion}
The $\pm J$ spin glass model in 2D can be used to efficiently study
the thermodynamics of a glassy model. Large scale simulations verify conceptually
novel scaling arguments for this model: these concepts have broad implications.
Strictly correlated spins at $T=0$ form fractal clusters that
can be explained using a droplet theory \cite{hartmann}, leading to backbone
correlations that decay as a power law.
Yet the spin-spin correlations at $T=0$ have long range order.
Though large zero-energy excitations are abundant,
the scale-dependent entropy of these excitations prevents them
from destabilizing the long range order.
This entropic ``stiffness'' at $T=0$ is very similar to the (free-)energetic stiffness in $d\ge 3$
models that gives a stable spin glass with long range order at nonzero $T$, making the 2D
$\pm J$ model a good candidate for exploring the spin glass phase in general.
The difference between discrete and continuous disorder is a general problem in statistical
mechanics; here we show how this difference can have striking effects on the scaling through a temperature dependent
crossover length.
When the temperature is nonzero,
application of this crossover length $\ell_{x}(T)$ leads to a new prediction
for the low temperature specific heat of the $\pm J$ model in 2D, which we have
verified numerically.

This work was supported in part by NSF grants DMR-1006731 and DMR-0819860. We are
grateful for the use of otherwise idle time on the Syracuse University
Gravitation and Relativity computing cluster, supported in part by NSF
grant PHY-0600953, and the Brutus cluster of ETH Zurich.

\end{document}